\documentclass[12pt]{article}

\usepackage{graphicx}
\usepackage{amssymb}
\usepackage{epstopdf}
\usepackage{color}
\usepackage{float}
\usepackage{ulem}
\usepackage{graphicx}
\usepackage{pdfpages}
\usepackage{amsmath}
\usepackage{listings}
\usepackage{amsmath}
\usepackage{amsthm}
\usepackage{mathrsfs}
\usepackage{amsfonts}
\usepackage{latexsym}
 
\usepackage[colorlinks,linkcolor=blue,urlcolor=blue,citecolor=blue]{hyperref}

\usepackage{cleveref}
\usepackage{appendix}
\usepackage{mathrsfs}  
\usepackage[utf8]{inputenc}

\newlength{\TZ}
\def\be{\begin{equation}}
\def\ee{\end{equation}}
\def\bea{\begin{eqnarray}}
\def\eea{\end{eqnarray}}

\newcommand{\D}{{\rm d}}                

\catcode`\@=11
\def\numberbysection{\@addtoreset{equation}{section}
        \def\theequation{\thesection.\arabic{equation}}}

\usepackage{xcolor}
\definecolor{dgreen}{rgb}{0,0.7,0}

\usepackage{cancel}

\begin{document}

\begin{titlepage}

\vskip 1.5 cm
\begin{center}
{\LARGE \bf Heterogeneous diffusion with recruitment and mortality}
\end{center}

\vskip 1.5 cm
\centerline{\bf{Esther D. Guti\'errez}$^{a}$, {\bf Alexander L\'opez}$^{a}$,  {\bf R. K. Singh}$^{b}$,}
\centerline{{\bf S\'ebastien Fumeron}$^{c}$, 
{\bf Malte Henkel}$^{c,d}$ and {\bf Trifce Sandev}$^{e,f,g}$}

\vskip 0.5 cm
\centerline{$^a$Escuela Superior Polit\'ecnica del Litoral, ESPOL, Departamento de F\'isica,} \centerline{Facultad de Ciencias Naturales y Matem\'aticas, Campus Gustavo Galindo
 Km. 30.5 V\'ia Perimetral,}
 \centerline{P. O. Box 09-01-5863, Guayaquil, Ecuador}
\vspace{0.45cm}
\centerline{$^b$Department of Physics, Banaras Hindu University, Varanasi 221005, India}
\vspace{0.45cm}
\centerline{$^c$Laboratoire de Physique et Chimie Th\'eoriques (CNRS UMR 7019), Universit\'e de Lorraine Nancy,}
\centerline{B.P. 70239, F -- 54506 Vand{\oe}uvre l\`es Nancy Cedex, France}
\vspace{0.45cm}
\centerline{$^d$Centro de F\'{i}sica Te\'{o}rica e Computacional, Universidade de Lisboa,}
\centerline{Campo Grande, P -- 1749-016 Lisboa, Portugal}
\vspace{0.45cm}
\centerline{$^e$Research Center for Computer Science and Information Technologies,}
\centerline{Macedonian Academy of Sciences and Arts,}
\centerline{Bul. Krste Misirkov 2, MK -- 1000 Skopje, Macedonia}
\vspace{0.45cm}
\centerline{$^f$Institute of Physics, Faculty of Natural Sciences and Mathematics,
Ss. Cyril and Methodius University,}
\centerline{Arhimedova 3, MK - 1000 Skopje, Macedonia}
\vspace{0.45cm}
\centerline{$^g$Department of Physics, Korea University, Seoul 02841, Korea}
\vspace{0.5cm}

\begin{abstract}
We study a heterogeneous diffusion process with a position-dependent diffusion coefficient in the presence of independent recruitment and mortality. This framework generalizes heterogeneous diffusion under stochastic resetting, which is recovered as the special case of equal recruitment and mortality rates. At long times, the system approaches a non-equilibrium stationary state with a mean squared displacement that saturates to a value independent of the recruitment rate. Crucially, this saturation limit directly encodes the underlying transport regime via an effective exponent applied to the diffusion coefficient, and we characterize the non-trivial transition to this stationary regime. We further derive the mean first-passage time to a target and show that, for a fixed recruitment-to-mortality ratio, an optimal mortality rate minimizes the search time. This "mortality-assisted search" effect is confirmed by numerical evaluation. Reinterpreting recruitment and mortality as infection and removal processes, we apply the model to epidemic spreading and show that a moderate removal rate can counter-intuitively accelerate, rather than delay, the arrival of infection at a new spatial region. Beyond epidemiology, our framework offers broad relevance to systems featuring concurrent particle creation and destruction, such as bacterial growth, tracer diffusion in reactive fluids, or cellular population dynamics in crowded environments.  
\end{abstract}

\end{titlepage}

\section{\label{sec:level0} Introduction} Since the seminal work on stochastic search problem by Rastrigin~\cite{00}, stochastic search strategies have been found to be particularly suited for medium optimization. Recent formulations of models used to describe search strategies, include the Heterogeneous diffusion processes (HDPs), with space-dependent diffusion coefficient, as they can be  relevant for tracer-particle diffusion in biological cells or heterogeneous complex fluids~\cite{01}. They are also found in a number of real-world systems, such as diffusive macromolecules or sub-micron tracers in biological cells, as they account for  transport processes in media where the diffusion coefficient varies across space~\cite{02}. These HDPs generalize anomalous diffusion processes~\cite{03,04,05,06,report,08,09,010,011,012,013}, which makes them adequate to describe transport processes in media where the diffusion coefficient varies across space, and thus, the  particles move at different speeds depending on their exact physical location. This spatial variation leads to non-standard or anomalous diffusion dynamics, frequently observed in crowded biological cells, porous geological formations, and complex fluids.  Recently, position-dependent diffusion coefficient has been used in mesoscopic approaches to transport in heterogeneous and porous media~\cite{app1}, to describe transport processes on random fractals~\cite{app6}, in turbulent diffusion and turbulent flows~\cite{app2,app3}. Moreover, different socioeconomic models, such as noisy voter model for opinion dynamics, can be interpreted as heterogeneous diffusion processes~\cite{app4,app5}.

\noindent The role of HDPs with a conserved number of particles has been extensively investigated. However, many phenomena of practical interest — infectious spreading being a prime example — inherently involve open populations, where newly infected individuals enter the system and deceased or recovered individuals leave it, so that the instantaneous number of active particles fluctuates rather than remains fixed. Capturing this requires going beyond the conserved-particle framework. The aim of this paper is to extend our understanding of heterogeneous diffusion to such open stochastic systems, by introducing independent recruitment and mortality processes into the HDP framework. This generalizes the previously studied case of heterogeneous diffusion under stochastic resetting, which is recovered here as the special case of equal recruitment and mortality rates, and in which the number of particles remains conserved.

\noindent The paper is structured as follows. In Section 2 we introduce the model and summarize the main quantities of interest for the underlying heterogeneous diffusion process. In Section 3 we solve the HDP with recruitment and mortality, deriving the propagator, its normalization, and the associated moments, and we characterize the stationary state reached at long times. In Section 4 we turn to the first-passage properties of the process, deriving the survival probability and mean first-passage time, and showing that an optimal mortality rate exists that minimizes the time to first arrival. In Section 5 we apply these results to a model of epidemic spreading, interpreting recruitment and mortality as infection and removal processes and discussing the epidemiological implications of our findings. We conclude with a summary and discussion in Section 6, while Appendices A and B provide the detailed derivations of the propagator and first-passage quantities, respectively.
\section{\label{sec:level1}Model}
The diffusion process in a heterogeneous and disordered environment often becomes anomalous, i.e., the corresponding mean squared displacement (MSD) has a power-law dependence on time, i.e., $\left\langle x^{2}(t)\right\rangle \sim t^\beta$ ($\beta\neq1$), due to the crowded and bounded environment, constrained motion, variation of the local diffusion coefficient in space and/or time, the memory effects of the environment, etc.~\cite{report,report2,chaos_rev}. We distinguish subdiffusion for $0<\beta<1$, normal diffusion for $\beta=1$ and super-diffusion for $\beta>1$~\cite{report}. 

The generic heterogeneous diffusion process (HDP) in a system with one-dimensional spatial degree of freedom, assuming a position-dependent diffusion coefficient, can be defined by the following Langevin equation~\cite{Leibovich}
\begin{align}\label{le str}
    \dot{x}(t)=\sqrt{2\mathcal{D}(x)}\,\xi(t),
\end{align}
where $\xi(t)$ is white Gaussian noise with $\left\langle \xi(t)\right\rangle = 0, 
\left\langle \xi(t)\xi(t^{\prime})\right\rangle = \delta(t-t^{\prime})$. Since the noise here is multiplicative, depending on the integration of the equation one can consider It\^{o}~\cite{1,2}, Stratonovich ~\cite{stratonovich} or H\"anggi-Klimontovich ~\cite{4,5} interpretation of the noise. 
Such process shows anomalous diffusive behaviour. Frequently used models are the power-law, exponential, or logarithmic space-dependent diffusion coefficient. In this work, we choose a power-law dependent diffusion coefficient given as (see for example Ref.~\cite{chchme}),
\begin{align}\label{diff coeff}
    \mathcal{D}(x)=\mathcal{D}_{\alpha}|x|^{\alpha}, \quad \alpha<2,
\end{align}
with $\mathcal{D}_{\alpha}$ playing the role of an effective diffusion parameter.
The resulting MSD for such HDPs behaves as \cite{chchme} \begin{align}
    \left\langle x^{2}(t)\right\rangle \sim t^{p}, \quad p=\frac{2}{2-\alpha},
\end{align}
which exhibits different diffusion regimes. Particularly, for $\alpha=0$ ($p=1$), we have normal diffusion. The case of sub-diffusion is observed for $\alpha<0$ ($0<p<1$), while super-diffusion for $0<\alpha<1$ ($1<p<2$). Additionally, for $1<\alpha<2$ ($p>2$) one observes hyper-diffusion, and the result for $\alpha=1$ corresponds to the ballistic motion ($p=2$). 

For the Stratonovich interpretation of the stochastic integrals, the corresponding Fokker-Planck equation for the probability density function (PDF) $P(x,t)$ reads~\cite{chchme,stratonovich,Leibovich}
\begin{align}\label{diffusion eq heterogeneous} 
\frac{\partial}{\partial t}P_0(x,t)=\mathcal{D}_\alpha\frac{\partial}{\partial x}\left\{|x|^{\alpha/2}\frac{\partial}{\partial x}\left[|x|^{\alpha/2}P_0(x,t)\right]\right\}.
\end{align}
The initial condition is chosen as $P_0(x,t=0)=\delta(x-x_0)$, while the boundary conditions are set to zero at infinity. The PDF of the HDP without recruitment and mortality reads~\cite{jpa_hdp}
\begin{align}\label{PDF str final}
    P_0(x,t)=\frac{|x|^{1/p-1}}{\sqrt{4\pi\mathcal{D}_{\alpha}t}}\exp\left(-\frac{p^{2}\left[\textrm{sgn}(x)\,|x|^{1/p}-\textrm{sgn}(x_0)\,|x_0|^{1/p}\right]^{2}}{4\mathcal{D}_{\alpha}t}\right).
\end{align}

\section{HDP with recruitment and mortality}\label{sec_no_reset}

HDPs have been extensively investigated in the literature, including HDPs with resetting~\cite{jpa_hdp}. It has been observed that in case of Poissonian resetting, the system approaches a non-equilibrium stationary state (NESS) in the long time limit~\cite{jpa_hdp}, which is different from the Laplace distribution observed for the standard diffusion process under resetting~\cite{evans_maj,evans_rev}. Here, we consider a more general model by considering that the number of trajectories returned to the initial state is different from the number of generated trajectories from the initial state. Therefore, we consider the following equation
\begin{align}\label{diffusion eq heterogeneous rm} 
\frac{\partial}{\partial t}P(x,t)=\mathcal{D}_\alpha\frac{\partial}{\partial x}\left\{|x|^{\alpha/2}\frac{\partial}{\partial x}\left[|x|^{\alpha/2}P(x,t)\right]\right\}-\mu P(x,t)+\lambda\delta(x-x_0),
\end{align}
where the second and third terms, from the right hand side of the equation, are the loss of density due to mortality at $x$ and the gain of density due to recruitment at $x_0$, respectively. For $\mu=\lambda=r$ we recover the corresponding model under stochastic resetting~\cite{jpa_hdp}. Here $\lambda$ and $\mu$ are recruitment and mortality rates, respectively. The analogous problem of standard diffusion with mortality and recruitment, which can be recovered for $\alpha=0$, was recently studied in Refs.~\cite{linn2026dynamic,mercado2026stochastic}, while the special case with $\alpha=2$ and $x>0$, which corresponds to the geometric Brownian motion under recruitment and mortality was recently investigated in Ref.~\cite{arxiv_gbm}. In Appendix~\ref{appa}, we show that the solution to Eq.~\eqref{diffusion eq heterogeneous rm} is
\begin{equation}\label{renewal rm}
    P(x,t)=e^{-\mu t}P_0(x,t)+\lambda\int_{0}^{t}e^{-\mu t'}P_0(x,t')\,\D t'\,,
\end{equation}
where $P_0(x,t)$ is given in Eq.~\eqref{PDF str final}, corresponding to $\mu=\lambda=0$. Despite the fact that the emerging properties of the modified system can be generated from the standard HDP, there are qualitative differences among them. For instance, $P(x,t)$ is in general non conservative, as we find (see Appendix~\ref{appa} for details),
\begin{align}
    \int_{-\infty}^{\infty}P(x,t)\,\D x&=1-\frac{\mu-\lambda}{\mu}\left(1-e^{-\mu t}\right)=N(t),
\end{align}
where we use the normalization of the PDF for the HDP, $\int_{-\infty}^{\infty}P_0(x,t)\,\D x\,=1$ (note that for $t\rightarrow\infty$, $N(t)=\lambda/\mu$). This encodes the generic asymmetry between recruitment and mortality rates. Thus, the normalized solution is expressed as
\begin{align}
    P_{N}(x,t)=\frac{1}{N(t)}P(x,t),
\end{align}
from which the MSD for $x_0=0$ is found to read as
\begin{align}{\label{msdx0}}
    \langle x^2(t)\rangle&=\frac{1}{N(t)}\frac{\Gamma\left(1+2p\right)}{p^{2p}}\frac{\mathcal{D}_\alpha^{p}}{\Gamma\left(1+p\right)}\left[e^{-\mu t}t^p+\lambda\int_0^{t}e^{-\mu t'}{t'}^p\,\D t'\,\right]\nonumber\\&=\frac{1}{N(t)}\frac{\Gamma\left(1+2p\right)}{p^{2p}}\frac{\mathcal{D}_\alpha^{p}}{\Gamma\left(1+p\right)}\left[e^{-\mu t}t^p+\lambda \frac{\gamma(p+1,\mu t)}{\mu^{p+1}}\right],
\end{align}
where $\gamma(a,z)=\int_{0}^{z}y^{a-1}e^{-y}\,dy$ is the lower incomplete gamma function. The  higher order moments and $x_0\neq0$ can be expressed in terms of the moments obtained from $P_0(x,t)$, and are given in Appendix~\ref{appa}. In the long-time limit $t\rightarrow\infty$, the MSD given in Eq.~\eqref{msdx0} saturates to
\begin{align}
    \langle x^2\rangle_{\text{sat}}\sim \frac{\Gamma\left(1+2p\right)}{p^{2p}}\frac{\mathcal{D}_\alpha^{p}}{\mu^{p}}.
\end{align}
This can be understood by considering that at long times, the average area explored by the particle cloud reaches a fixed saturation limit that is completely unaffected by how fast new particles are injected. Physically, the recruitment rate simply dictates the total number of particles present, scaling the overall density of the cloud without altering its geometric shape, causing it to cancel out during statistical normalization. The final spatial boundary is instead governed by a dynamic balance between transport efficiency and particle lifespans, where mortality acts as a natural cut-off preventing infinite migration. The model’s broad relevance stems from how this saturation limit directly captures the underlying transport regime; the anomalous diffusion coefficient is raised to an effective parameter that shifts predictably depending on whether the system is subdiffusive or superdiffusive. The saturated deviation thus serves as a sensitive diagnostic tool, revealing the precise physical nature of the medium regardless of injection intensity.
\section{First-passage time}
We now consider the first-passage properties of the heterogeneous diffusion process in the presence of recruitment and mortality, following the approach given in references.~\cite{linn2026dynamic,mercado2026stochastic,arxiv_gbm}. Let the target be located at $x=L$, and assume that trajectories are initially injected at $x_0$. To define the first-passage problem, we impose an absorbing boundary condition at the target,
\begin{equation}
    P(L,t)=0.
\end{equation}
This condition means that once a trajectory reaches $x=L$, it is removed from the process.
For the underlying heterogeneous diffusion process, in the absence of both recruitment and mortality, we denote it as $   T_0=\inf\{t>0:X(t)=L\},$
and the corresponding survival probability and first-passage time density are given by
\begin{equation}
    S_0(t)=\mathbb{P}(T_0>t),
    \qquad
    f_0(t)=-\frac{\D S_0(t)}{\D t}.
    \label{FPT_base_survival}
\end{equation}
Here, $S_0(t)$ is the probability that a single trajectory of the underlying HDP has not reached the target up to time $t$, and $f_0(t)$ is its first-passage time density. In Appendix~\ref{appb} we show explicitly that the free survival probability in absence of mortality and recruitment is given by
\begin{equation}\label{S0}
    S_0(t)
    =
    \operatorname{erf}
    \left(
    \frac{|y_L-y_0|}{\sqrt{4\mathcal{D}_\alpha t}}
    \right),
\end{equation}
where we use the effective coordinate $y (x)=p\,\operatorname{sgn}(x)|x|^{1/p},$ while $\text{erf}(x)$ is the error function.
    
\noindent When mortality is included, but recruitment is still absent, a trajectory contributes to a successful first-passage event only if it reaches the target before being removed. Since mortality occurs at rate $\mu$, the probability that a trajectory survives up to time $t$ is $e^{-\mu t}$. Therefore, the density of successful arrivals at the target is
    $e^{-\mu t}f_0(t)$.
The total probability that a single trajectory reaches the target before being removed is then
\begin{equation}
    \int_0^\infty
    e^{-\mu t}f_0(t)\,\D t\,
    =
    \hat f_0(\mu),
    \label{FPT_success_probability}
\end{equation}
where $\hat f_0(s)$ denotes the Laplace transform of the first-passage time density of the underlying HDP.
Equivalently, the survival probability of a single mortal trajectory, without recruitment, is
\begin{equation}
    S_{\mu}(t)
    =
    \mathbb{P}(T_{\mu}>t)
    =
    1-
    \int_0^t
    e^{-\mu t'}f_0(t')\,\D t'\,.
    \label{FPT_single_mortal_survival}
\end{equation}

This quantity represents the probability that the trajectory has not reached the target up to time $t$, including the possibility that it has been removed by mortality before reaching the target.

In the presence of recruitment, the first-passage event corresponds to the first arrival of any trajectory injected into the system. Thus, the relevant survival probability is not the survival probability of a single trajectory, but rather the probability that no recruited trajectory has reached the target up to time $t$. Following the dynamic redundancy and mortality framework of Ref.~\cite{linn2026dynamic}, we define the survival probability of the full recruitment--mortality process as $S(t)\equiv S_{\lambda,\mu}(t)$

\begin{equation}
     S(t)
    =
    \mathbb{P}(T>t),
\end{equation}
where $T$ denotes the first time at which any trajectory in the recruited population reaches the target.
For dynamically recruited and mortal trajectories, the full survival probability can be written as (see Appendix~\ref{appb})
\begin{equation}
    S(t)=
    S_{\mu}(t)
    \exp\left[
    -\lambda
    \int_0^t
    \left(
    1-S_{\mu}(t')
    \right)\D t'\,
    \right].
    \label{FPT_full_DRM_survival}
\end{equation}
This formulation is preferable to obtaining the first-passage density from the flux of the mean density $P(x,t)$, because in the recruitment--mortality process the number of trajectories is not conserved: new trajectories are injected at $x_0$ with rate $\lambda$, while existing trajectories are removed with rate $\mu$.
The first-passage time density of the full process is then
\begin{equation}
   f (t)=
    -\frac{\D S(t)}{\D t}\label{FPT_full_density}.
\end{equation}
Being a stochastic process, the relevant quantity of interest is the average first-passage time $\langle T\rangle$, which is obtained as 
\begin{equation}
    \langle T\rangle=
    \int_0^\infty
    S(t)\D t\label{FPT_full_MFPT}
\end{equation}

Therefore, the first-passage problem separates into two steps. First, one must determine the first-passage density $f_0(t)$ of a single heterogeneous diffusion trajectory without mortality and recruitment. Second, this quantity is used to construct the mortal survival probability $S_{\mu}(t)$ and, subsequently, the survival probability of the full recruitment--mortality process.
 In Appendix~\ref{appb}, it is shown that the survival probability of a single mortal HDP trajectory without recruitment is given by
\begin{align}
    S_{\mu}(t)
    =
    1
    -
    \frac{1}{2}
    \exp\left[
    -\Delta y\sqrt{\frac{\mu}{\mathcal{D}_\alpha}}
    \right]
    \operatorname{erfc}
    \left(
    \frac{\Delta y}{\sqrt{4\mathcal{D}_\alpha t}}
    -
    \sqrt{\mu t}
    \right) -
    \frac{1}{2}
    \exp\left[
    \Delta y\sqrt{\frac{\mu}{\mathcal{D}_\alpha}}
    \right]
    \operatorname{erfc}
    \left(
    \frac{\Delta y}{\sqrt{4\mathcal{D}_\alpha t}}
    +
    \sqrt{\mu t}
    \right),
    \label{survival_exact}
\end{align}
where we have used the effective coordinate $y(x)$, such that the effective distance is 

\begin{equation*}
\Delta y 
=
p\left|
\operatorname{sgn}(L)|L|^{1/p}
-
\operatorname{sgn}(x_0)|x_0|^{1/p}
\right|.
\end{equation*}

If we restrict the analysis to the positive axis, with $0<x_0<L$, this expression reduces to 

\begin{equation}
    \Delta y
    =
    \frac{2}{2-\alpha}
    \left(
    L^{(2-\alpha)/2}
    -
    x_0^{(2-\alpha)/2}
    \right),
    \label{effective_distance_positive_axis}
\end{equation}
while $\text{erfc}(x)=1-\text{erf}(x)$ is the complementary error function.
Once the single-trajectory  survival probability 
$S_{\mu}(t)$ has been obtained, recruitment can be incorporated through the dynamic redundancy formulation. The survival probability of the full recruitment--mortality process is given by Eq.~\eqref{FPT_full_DRM_survival}. Because the explicit expression for $S_{\mu}(t)$ involves complementary error functions, the resulting time integral in Eq.~\eqref{FPT_full_DRM_survival} does not lead to a compact or physically transparent closed form, (see Appendix~\ref{appb}). We therefore evaluate the recruitment contribution numerically. 

Before analysing  results obtained from the numerical evaluation, it is useful to examine the long-time behaviour of Eq.~\eqref{FPT_full_DRM_survival}. From Eq.~\eqref{survival_exact}, the probability that a single mortal trajectory has reached the target before time $t$ satisfies
\begin{equation}
    1-S_{\mu}(t)
    \longrightarrow \exp\left[
    -\Delta y
    \sqrt{\frac{\mu}{\mathcal{D}_\alpha}}
    \right]\equiv
    \mathcal{P}_{\mu},
    \qquad
    t\to\infty\label{22},
\end{equation}
\noindent Therefore, for sufficiently long times,
\begin{equation}
    \int_0^t
    \left[
    1-S_{\mu}(t')
    \right]\D t'\,
    \sim   \mathcal{P}_{\mu}t .
\end{equation}
Substitution into Eq.~\eqref{FPT_full_DRM_survival} gives the asymptotic behaviour
\begin{equation}
    S(t)\sim
    S_{\mu}(\infty)
    \exp\left[
    -\lambda \mathcal{P}_{\mu}t
    \right],
    \qquad
    t\to\infty.
    \label{full_survival_long_time}
\end{equation}
Thus, recruitment induces an effective exponential decay at long times, with rate
\begin{equation}
    \lambda_{\mathrm{eff}}
    =
    \lambda\mathcal{P}_{\mu}.
    \label{lambda_effective}
\end{equation}

\begin{figure}[htbp]
    \centering
    \includegraphics[width=0.8\textwidth]{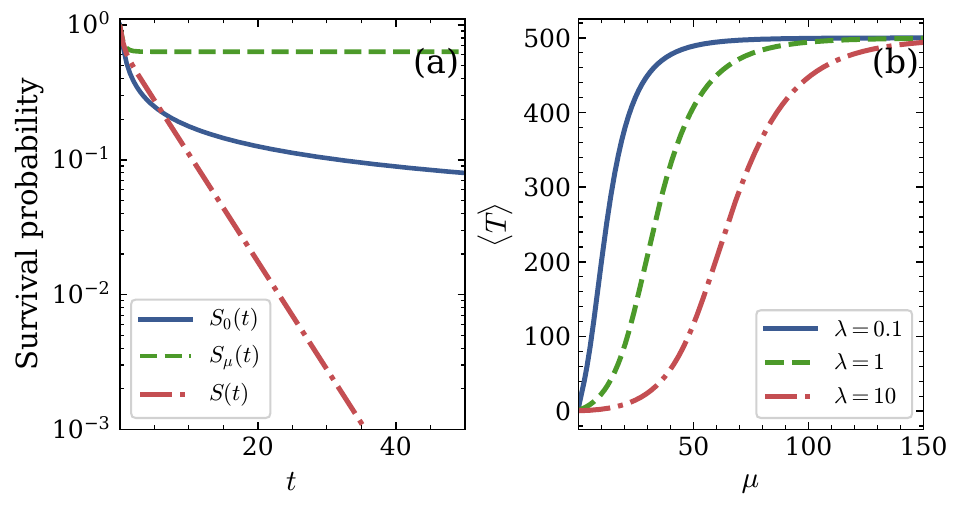}
    \caption{
    Numerical evaluation of the first-passage properties of the heterogeneous diffusion process with mortality and recruitment, in dimensionless units.
    (a) Survival probability of the underlying HDP, $S_0(t)$; of a single mortal trajectory without recruitment, $S_{\mu}(t)$ [Eq.~\eqref{survival_exact}]; and of the full recruitment--mortality process, $S(t)$ [Eq.~\eqref{FPT_full_DRM_survival}], for fixed mortality and recruitment rates.
    (b) Numerical mean first-passage time $\langle T\rangle$ as a function of the mortality rate $\mu$, for three fixed values of the recruitment rate $\lambda$.
    The three curves represent the regimes $\mu>\lambda$, $\mu=\lambda$, and $\mu<\lambda$ when compared with the reference value $\mu=1.0$ used in panel (a).
    The mean first-passage time in panel (b) was evaluated over the fixed numerical time window $0<t<500$.
    Parameters: $\Delta y=1$ and $\mathcal{D}_\alpha=1$.
    In panel (a), $\mu=1.0$ and $\lambda=0.5$.
    In panel (b), $\lambda=0.1, 1.0, 10.0$.
    }
    \label{fig_FPT_recruitment_mortality}
\end{figure}
This result provides a useful consistency check for the numerical evaluation: the tail of the full survival probability should decay exponentially with slope controlled by $\lambda\mathcal{P}_{\mu}$. In Fig.~\ref{fig_FPT_recruitment_mortality}, we show the numerical evaluation of the first-passage properties of the heterogeneous
diffusion process with mortality and recruitment, in dimensionless units. In panel (a), we show the survival probabilities without recruitment or mortality $S_0(t)$, the case of finite mortality  and vanishing recruitment $S_\mu(t)$ [Eq.~\eqref{survival_exact})], and $S(t)$ [Eq.~\eqref{FPT_full_DRM_survival}] when both mortality and recruitment are finite. We notice that $S_\mu(t)$ represents the probability that the trajectory has not reached the target up to time $t$, and therefore including trajectories that have already been removed by mortality before reaching it — such trajectories can never subsequently arrive, and are therefore counted as ``surviving" in the first-passage sense for all later times. This is the origin of a feature that may otherwise appear counter-intuitive: since one-dimensional diffusion is recurrent, the mortality-free survival probability satisfies $S_0(t)\rightarrow0$ as $t\rightarrow\infty$, whereas $S_\mu(t)$
 saturates to a non-zero plateau [Eq.~\eqref{survival_exact}] because a finite fraction of trajectories are permanently removed before arrival. As a result, $S_\mu(t)\ge S_0(t)$
at long times, even though mortality does not aid (and indeed only competes with) arrival at the target for any individual trajectory. The inclusion of recruitment reverses the plateau behaviour: while a single mortal trajectory has a finite probability of never reaching the target, and hence $S_\mu(t)$ 
saturates at long times, the full recruitment-mortality survival probability $S(t)$ counts the probability that none of the trajectories injected up to time $t$ has yet arrived. Each recruited trajectory constitutes an independent additional attempt, so $S(t)$ decays to zero even though each individual attempt is governed by the same $S_\mu(t)$ that itself saturates the population-level probability of ``no success yet", which is driven down by the growing number of parallel attempts, not by any change in single trajectory behaviour. This quantitative dependence on $\lambda$ is illustrated further in Fig.~\ref{fig_recruitment_survival_appendix} of Appendix~\ref{appb}, which shows that increasing the recruitment rate over two orders of magnitude drives correspondingly faster decay of  $S(t)$, consistent with the dynamic-redundancy interpretation above. 
Fig.~\ref{fig_FPT_recruitment_mortality}(b) shows the mean first-passage time $\langle T\rangle$ as a function of the mortality rate $\mu$, for three fixed values of the recruitment rate $\lambda$. Each curve exhibits a well-defined minimum at small $\mu$: the mean first-passage time initially decreases as mortality increases from zero, reaches a minimum at some finite $\mu^*$, and then rises and saturates at large $\mu$. This non-monotonic behaviour reflects the competition between two effects. At small $\mu$, increasing mortality removes trajectories that have wandered far from the target and are unlikely to arrive, effectively pruning unproductive searchers and allowing more recently recruited trajectories closer (in expectation) to the target, to dominate the search; this accelerates the first arrival. At large $\mu$, however, mortality begins to remove trajectories faster than recruitment can replace them with viable searchers, so that arrival is increasingly foreclosed altogether and the mean FPT grows and eventually saturates. The existence of a genuine minimum for every curve in Fig. 1(b) confirms that this mortality-assisted search effect is robust across the range of recruitment rates considered, and motivates a systematic characterization of the optimal mortality rate as a function of the recruitment-to-mortality ratio.

\begin{figure}[htbp]
    \centering
    \includegraphics[width=1.0\textwidth]{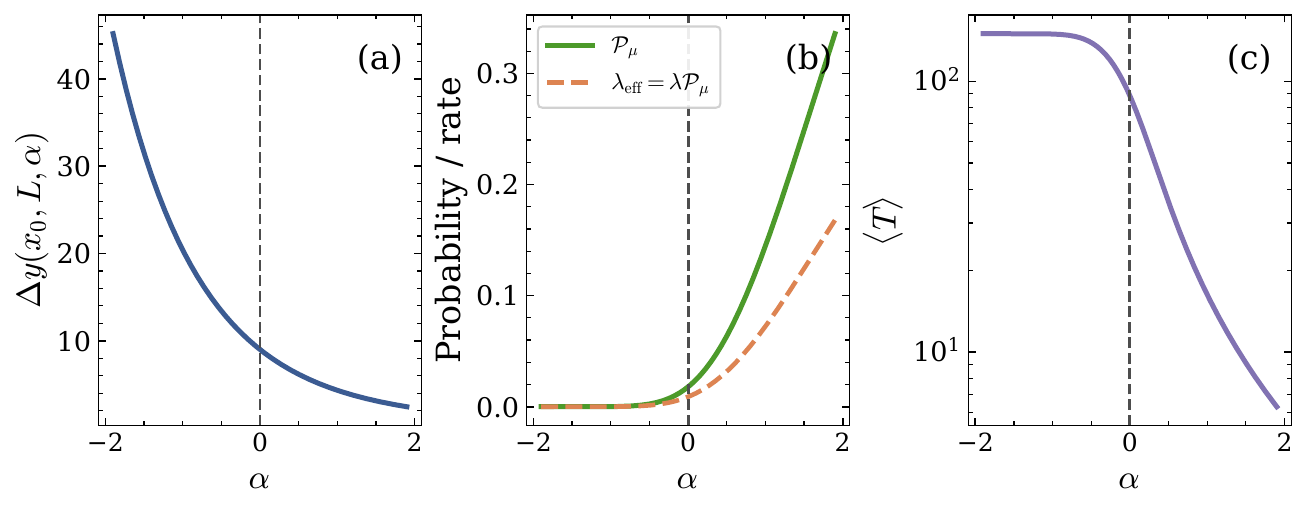}
\caption{
Effect of the heterogeneity exponent $\alpha$ on the first-passage properties, in dimensionless units. 
The physical initial position and target position are fixed at $x_0=1$ and $L=10$, so that the $\alpha$-dependence  enters entirely through the effective distance $\Delta y(x_0,L,\alpha)$ [Eq.\eqref{effective_distance_positive_axis}]. 
(a) $\Delta y$ decreases monotonically with $\alpha$, reflecting how larger heterogeneity makes the target dynamically closer to the starting point. 
(b) Single-trajectory success probability $\mathcal{P}_{\mu}$ [Eq.\eqref{22}] and the corresponding effective long-time decay rate $\lambda_{\mathrm{eff}}=\lambda\mathcal{P}_{\mu}$ [Eq.~\eqref{lambda_effective}], both increasing with $\alpha$ as the effective distance to the target shrinks. 
(c) Mean first-passage time $\langle T\rangle$ which shows a crossover from an almost $\alpha$-independent plateau at negative $\alpha$  to a monotonic decrease with $\alpha$, consistent with the shrinking effective distance shown in panel (a). 
Parameters: $\mathcal{D}_\alpha=1$, $\mu=0.2$, and $\lambda=0.5$. 
}
\label{fig_alpha_dependence}
\end{figure}

To make the role of the heterogeneity exponent explicit, we keep the physical positions $x_0$ and $L$ fixed and analyse how the effective distance in Eq.~\eqref{effective_distance_positive_axis} varies with $\alpha$. This is important because fixing $\Delta y$ masks the explicit dependence on the heterogeneous diffusion exponent. In dimensionless units, with $\mathcal{D}_\alpha=1$, the variation of $\Delta y(x_0,L,\alpha)$ directly determines the single-trajectory success probability in Eq.~\eqref{22} and, consequently, the effective long-time decay rate in Eq.~\eqref{lambda_effective}. Fig.~(\ref{fig_alpha_dependence}) illustrates this chain of dependencies explicitly, for fixed initial and target positions $x_0=1$ $L=10$, respectively. Panel (a) shows that the effective distance $\Delta y(x_0,L,\alpha)$
decreases monotonically with $\alpha$, such that larger heterogeneity exponents correspond to a diffusion coefficient that grows more steeply with $|x|$, so that trajectories move far from the origin increasingly faster, and the target becomes dynamically closer, even though its physical position $L$ is unchanged. Panel (b) shows the corresponding single-trajectory success probability $\mathcal{P}_\mu$ [Eq.\eqref{22}] and effective decay rate $\lambda_{eff}=\lambda\mathcal{P}_\mu$ [Eq.~\eqref{lambda_effective}], both of which increase with heterogeneity as a direct consequence of the shrinking effective distance in panel (a). Hence, a trajectory is more likely to reach the target before being removed by mortality when that target is effectively closer. Panel (c) shows the resulting mean first-passage time $\langle T\rangle$ of the full recruitment–mortality process, which exhibits a crossover from an almost $\alpha$-independent plateau at negative values of the heterogeneity parameter — where the target is effectively far and mortality dominates the outcome regardless of moderate changes in $\alpha$ — to a regime of monotonic decrease at less negative and positive $\alpha$, where the shrinking effective distance increasingly accelerates arrival. Taken together, Fig.~\ref{fig_alpha_dependence} demonstrates that the heterogeneity exponent $\alpha$, and not only the mortality and recruitment rates, is itself a key control parameter for the first-passage properties of the process, acting entirely through its effect on the effective distance $\Delta y(x_0,L,\alpha).$

\begin{figure}[htbp]
    \centering
    \includegraphics[width=1.0\textwidth]{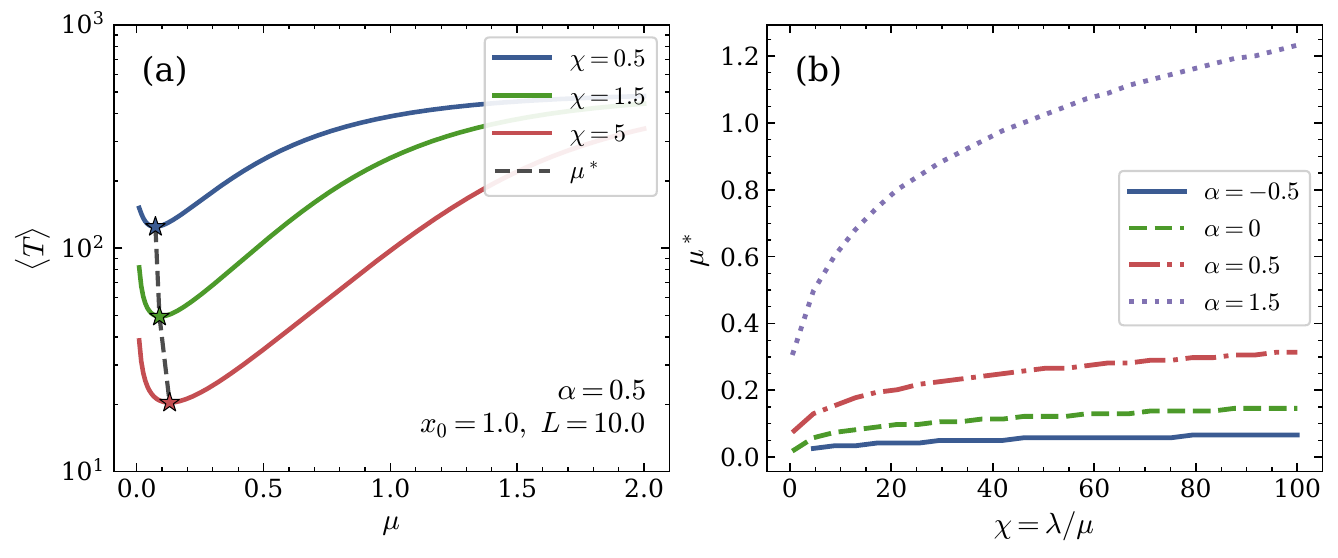}
\caption{Optimal mortality rate in the first-passage time properties of the heterogeneous diffusion process with recruitment and mortality, in dimensionless units, for fixed initial and target positions $x_0=1$, $L=10$. 
(a) Mean first-passage time $\langle T\rangle$ as a function of the mortality rate $\mu$, for different values of the recruitment--mortality ratio $\chi=\lambda/\mu$, at fixed heterogeneity exponent $\alpha=0.5$. For each curve, the recruitment rate is set as $\lambda=\chi\mu$. Every curve shows a well-defined minimum, confirming the existence of an optimal mortality rate  $\mu^*$ that minimizes the mean first-passage time; star markers indicate the numerically obtained minima, and the dashed curve traces the locus of $\mu^*$ as $\chi$ is varied. 
(b) Optimal mortality rate $\mu^*$ as a function of $\chi$, for different values of the heterogeneity exponent $\alpha$, whose effect enters through the effective distance $\Delta y(x_0,L,\alpha)$, and $\mathcal{D}_\alpha=1$. For all $\alpha$, $\mu^*$ increases with $\chi$ and saturates at large $\chi$; larger $\alpha$ systematically shifts $\mu^*$ to higher values, consistent with the shrinking effective distance to the target discussed in Fig. \ref{fig_alpha_dependence}. 
}
\label{fig_optimal_mortality_alpha}
\end{figure}

To compare the role of recruitment and mortality, we introduce the dimensionless ratio
\begin{equation}
    \chi=\frac{\lambda}{\mu},
\end{equation}
so that $\lambda=\chi\mu$. For fixed heterogeneity exponent $\alpha$, we then analyse the dependence of the mean first-passage time on the mortality rate,
which makes it  possible to identify whether an optimal mortality rate $\mu^*$ exists for each recruitment--mortality ratio $\chi$, as we can see in Fig.~\ref{fig_optimal_mortality_alpha}. The results shown in this figure confirm this systematically. Panel~(a) shows $\langle T\rangle$ as a function of $\mu$ for several fixed values of $\chi$, at fixed $\alpha=0.5$. Every curve exhibits a single well-defined minimum, generalizing the fixed $\lambda$ result of Fig.~\ref{fig_FPT_recruitment_mortality}(b) to the full range of recruitment-to-mortality ratios: for each $\chi$, there is a distinct optimal mortality rate $\mu^*$ that minimizes the mean first-passage time, and the locus of these minima (dashed curve, star markers) shifts to smaller $\mu$ as $\chi$ increases. This trend has a natural interpretation: as recruitment becomes relatively stronger (larger $\chi$), fewer removals per unit time are needed to prune unsuccessful trajectories and let recently recruited, better-positioned trajectories dominate the search, so the optimal mortality rate decreases. Panel~(b) characterizes this dependence directly, showing $\mu^*$ as a function of $\chi$ for several values of $\alpha$. For every heterogeneity value $\alpha$, $\mu^*$  increases with $\chi$ and saturates at large $\chi$, consistent with panel (a): as recruitment grows very strong relative to mortality, the optimal mortality rate approaches a constant value, since further increases in $\chi$ contribute diminishing additional benefit from pruning. Comparing curves across $\alpha$ shows that larger heterogeneity exponents shift $\mu^*$ to systematically higher values, in line with the shrinking effective distance to the target discussed in Fig.~\ref{fig_alpha_dependence}: since a more heterogeneous, effectively closer target is easier for trajectories to reach on their own, a higher mortality rate is required before pruning of unsuccessful trajectories becomes the dominant effect. Together, Figs.~\ref{fig_FPT_recruitment_mortality}–\ref{fig_optimal_mortality_alpha} establish that the mortality-assisted search effect is a robust feature of the heterogeneous diffusion process with recruitment and mortality, with the optimal mortality rate controlled jointly by the recruitment-to-mortality balance $\chi$ and the heterogeneity exponent $\alpha$ through the effective distance $\Delta y(x_0,L,\alpha)$.

\section{Application to epidemic spreading}\label{sec_bio}

The theoretical framework developed in the previous sections for heterogeneous diffusion with mortality and recruitment finds a natural and timely application in the modelling of epidemic spreading. The recent COVID-19 pandemic has highlighted the urgent need for a deeper understanding of the mechanisms underlying disease propagation, in particular the role of spatial heterogeneity and mobility. While compartmental models such as SIR, SEIR, and their variants have become standard tools in epidemiology, they typically assume spatially homogeneous mixing or, at best, simple diffusive motion on networks \cite{Forgacs2022, Chindra2025}. This simplifying assumption is known to neglect important spatial heterogeneities and variations in connectivity, which can significantly affect the dynamics of disease spread \cite{Siegenfeld2022}. To overcome these limitations, recent efforts have turned to network-based and spatial models that better capture the heterogeneity of real-world transmission structures \cite{Pokhrel2026}. However, these approaches often lack a continuum description that simultaneously accounts for heterogeneous diffusion, birth-death processes, and long-range resetting dynamics.

In reality, the mobility of individuals is often heterogeneous: some regions are more accessible than others, and the rate at which individuals move depends on the local environment, population density or infrastructure. Moreover, the number of infectious individuals may not be constant over time: infected individuals may die (mortality), while new infections continuously arise from contacts between individuals (recruitment into the infectious compartment). 

We consider a population of infected individuals moving in a one-dimensional heterogeneous environment. Each infected individual is treated as a random walker and undergoes a heterogeneous diffusion process described by the space-dependent diffusion coefficient $\mathcal{D}(x)=\mathcal{D}_\alpha |x|^\alpha$, with $\alpha\leq 2$. This choice represents situations where the mobility of infected individuals depends on the spatial coordinate: for $\alpha>0$, mobility increases with distance from the origin (e.g., urban centres with higher population density and connectivity); for $\alpha<0$, mobility decreases with distance (e.g., rural or remote areas with limited infrastructure). The case $\alpha=0$ recovers standard Brownian diffusion. Our epidemic spreading model also includes two competing processes:
\begin{enumerate}
\item \textit{Mortality}: Infected individuals may die from the disease at rate $\mu$. This corresponds to a loss term $-\mu P(x,t)$ in the density equation, where $P(x,t)$ is the probability density of finding an infected individual at position $x$ at time $t$. Importantly, mortality is spatially local: individuals die at their current position.
\item \textit{Recruitment}: New infections are generated through contacts between susceptible and infected individuals. In a simplified description, we assume that new infections occur at a fixed spatial location $x_0$ (e.g., a densely populated urban centre, a hospital, or a super-spreading location) at rate $\lambda$. This corresponds to a source term $+\lambda\delta(x-x_0)$ in the density equation. More generally, recruitment could occur at multiple locations, but the single-source case already captures the essential phenomenology.
\end{enumerate}

Under these assumptions, the resulting evolution equation is precisely (\ref{diffusion eq heterogeneous rm}) with the interpretation that $P(x,t)$ is the density of infected individuals (the total number of infected individuals is not conserved). The mortality rate $\mu$ and the recruitment rate $\lambda$ play the roles of the death rate and the infection rate, respectively. When $\lambda=\mu$, the system reduces to the standard stochastic resetting scenario, where each trajectory is reset to $x_0$ at rate $r$. The general solution (\ref{renewal rm}) has a clear epidemiological meaning. The first term represents infected individuals that were present at time $t=0$ and have survived (not died) up to time $t$, without any new infections. The second term represents the contribution of all infections that occurred at earlier times $t'$ (recruited at rate $\lambda$) and have survived until time $t$. Thus, the solution is a renewal equation: the current density of infected individuals is the sum of the surviving initial population and the surviving newly infected individuals.

The survival probability of the full recruitment--mortality process, given by Eq.~(\ref{FPT_full_DRM_survival}), has a clear epidemiological meaning as well: it corresponds to the probability that no infected individual has reached the target region up to time $t$. In epidemiology, a target region consists in a specific location that acts as as a primary hub for long-range transmission, such as airports with large flight volumes, highly-connected cities or most central cities in the mobility network. The notion of epidemiologically relevant nodes is supported by Bobashev et al. \cite{Bobashev2008}, who showed that a relatively small sample of about 200–300 cities was enough to capture the global spread dynamics of a disease such as influenza. The long-time decay of the full survival probability is given by Eq.~(\ref{full_survival_long_time}). This result is of prime importance for epidemic management: the rate at which the probability of no arrival decays is proportional to the product of the recruitment (infection) rate and the single-trajectory success probability. This explains why interventions that reduce either $\lambda$ (e.g., social distancing, lockdowns) or $\mathcal{P}_{\mu}$ (e.g., reducing mobility, increasing the removal rate $\mu$ through quarantine or treatment) will slow down the spread of the epidemic to target regions.

The mean first-passage time, $\langle T\rangle$, quantifies the expected time until the first infected individual reaches the target. Fig.~\ref{fig_optimal_mortality_alpha} shows that for each value of the recruitment--mortality ratio $\chi=\lambda/\mu$, there exists an optimal mortality rate $\mu^*$ that minimises the mean first-passage time. This is a striking result: a certain level of mortality can actually accelerate the arrival of the first infected individual to the target. The reason is that mortality removes infected individuals that are far from the target and unlikely to reach it, thereby effectively selecting for trajectories that are closer to the target. This counter-intuitive effect, known as the ``mortality-assisted search'' or ``dynamic redundancy'' phenomenon, has been recently identified in the context of stochastic search with a fluctuating number of searchers \cite{Meerson2015,Campos2024}. Our results extend this phenomenon to the epidemic context: a moderate death rate among infected individuals can paradoxically hasten the arrival of the infection to a new region, because it eliminates the ``unsuccessful'' trajectories and allows newly recruited individuals to start afresh from the recruitment point. This prediction could, in principle, be tested empirically by comparing outbreak dynamics in regions with different case fatality rates or different quarantine efficiencies.

\section{Conclusions}\label{sec_summary}

We have introduced and solved a model of heterogeneous diffusion in the presence of independent recruitment and mortality, generalizing the previously studied case of heterogeneous diffusion under stochastic resetting (recovered here as the special case $\lambda=\mu$). Using a Laplace-transform mapping between the recruitment–mortality problem and the underlying resetting-free HDP, we derived the corresponding mean squared displacement and showed that, remarkably, its long-time saturation value depends only on the mortality rate $\mu$, and the heterogeneity exponent $\alpha$, and is independent of the recruitment rate $\lambda$. Therefore, the diffusive spread of the population in steady state is set entirely by how fast individuals are removed, not by how fast they are replenished.
Physically, this independence arises because the replenishment rate acts as a linear scaling factor, adjusting the total number of individuals without altering the geometric shape of the population cloud. Crucially, the model's structural relevance is reflected in how this final saturation limit directly encodes the underlying transport regime, with the anomalous diffusion coefficient raised to an effective parameter that shifts predictably across the subdiffusive and superdiffusive spectrum.
We then analyzed the first-passage properties of the process to an absorbing target, combining the single-trajectory survival probability under mortality with a dynamic-redundancy description of recruitment. This yields a survival probability that decays exponentially at long times with an effective rate $\lambda_{eff}= \lambda\mathcal{P}_\mu$, where $\mathcal{P}_\mu$ is the probability that a single mortal trajectory reaches the target before being removed. Most notably, we found that the mean first-passage time $\langle T\rangle$ is a non-monotonic function of the mortality rate for fixed recruitment-to-mortality ratio $\chi=\lambda/\mu$, exhibiting a well-defined optimal mortality rate $\mu^*$ that minimizes the time to first arrival. This "mortality-assisted search" effect arises because mortality preferentially removes trajectories that are far from the target and unlikely ever to reach it, effectively pruning unsuccessful searchers and allowing freshly recruited trajectories, closer in expectation to the target, to dominate the search. We characterized how $\mu^*$ depends on both $\chi$ and the heterogeneity exponent $\alpha$ through the effective distance $\Delta y(x_0,L,\alpha)$, showing that stronger heterogeneity (larger $\alpha$) systematically raises the optimal mortality rate needed to achieve the fastest first arrival.
Reinterpreting the model in the context of epidemic spreading, we showed that recruitment and mortality map naturally onto infection and death/removal processes, and that the mean first-passage time corresponds to the expected arrival time of the first infected individual at a spatially remote target region. Our results suggest the counter-intuitive prediction that, within this framework, a moderate case-fatality or removal rate can accelerate rather than delay the arrival of infection at a new region; a phenomenon with a clear resetting-search analog but, to our knowledge, not previously identified in a continuum spatial epidemic model with heterogeneous mobility. This suggests that interventions altering the effective removal rate (through quarantine, treatment, or case detection) may have non-trivial, non-monotonic effects on the spatial spreading time of an outbreak, distinct from their effect on the total case count.
Natural extensions of this work include higher-dimensional generalizations, multiple recruitment sources, time-dependent or state-dependent recruitment and mortality rates (e.g., mortality that itself depends on time since infection), and a more detailed comparison with agent-based or network epidemic models to test the robustness of the mortality-assisted search effect beyond the mean-field, single-source setting considered here.

\noindent{\bf Acknowledgments:} {RKS acknowledges financial support from Anusandhan National Research Foundation via the Ramanujan Fellowship project RJF/2026/000006. TS is supported by Korea University and the Alliance of International Science Organizations (Project No. ANSO-CR-PP2022-05). 
TS also acknowledges support and hospitality by Dragi Karevski and the Laboratory of Theoretical Physics and Chemistry, University of Lorraine in Nancy.}

\appendix
\section{Solution of HDP with recruitment and mortality}\label{appa}
In this section we  give the derivations to solve the equation for the heterogeneous diffusion equation with recruitment and mortality
\begin{align}\label{hdprm} 
\frac{\partial}{\partial t}P(x,t)=\mathcal{D}_\alpha\frac{\partial}{\partial x}\left\{|x|^{\alpha/2}\frac{\partial}{\partial x}\left[|x|^{\alpha/2}P(x,t)\right]\right\}-\mu P(x,t)+\lambda\delta(x-x_0),
\end{align}
To this end, we first Laplace transform in time, to get
\begin{align}\label{diffusion eq heterogeneous rm l} 
s\hat{P}(x,s)-\delta(x-x_0)=\mathcal{D}_\alpha\frac{\partial}{\partial x}\left\{|x|^{\alpha/2}\frac{\partial}{\partial x}\left[|x|^{\alpha/2}\hat{P}(x,s)\right]\right\}-\mu\hat{P}(x,s)+\frac{\lambda}{s}\delta(x-x_0),
\end{align}
which can be rewritten as
\begin{align}\label{diffusion eq heterogeneous rm l2} 
(s+\mu)\hat{P}(x,s)-\frac{s+\lambda}{s}\delta(x-x_0)=\mathcal{D}_\alpha\frac{\partial}{\partial x}\left\{|x|^{\alpha/2}\frac{\partial}{\partial x}\left[|x|^{\alpha/2}\hat{P}(x,s)\right]\right\}.
\end{align}
Contrary to this, by the Laplace transformation of Eq.~(\ref{diffusion eq heterogeneous}) in the main text (i.e., without recruitment and mortality), we have
\begin{align}\label{diffusion eq heterogeneous l} 
s\hat{P}_0(x,s)-\delta(x-x_0)=\mathcal{D}_\alpha\frac{\partial}{\partial x}\left\{|x|^{\alpha/2}\frac{\partial}{\partial x}\left[|x|^{\alpha/2}\hat{P}_0(x,s)\right]\right\}.
\end{align}
Let us exchange $s\rightarrow s+\mu$ in eq.~(\ref{diffusion eq heterogeneous l}) and find
\begin{align}\label{diffusion eq heterogeneous l2} 
(s+\mu)\hat{P}_0(x,s+\mu)-\delta(x-x_0)=\mathcal{D}_\alpha\frac{\partial}{\partial x}\left\{|x|^{\alpha/2}\frac{\partial}{\partial x}\left[|x|^{\alpha/2}\hat{P}_0(x,s+\mu)\right]\right\}.
\end{align}
We multiply this equation from both sides by $(s+\lambda)/s$,
\begin{align}\label{diffusion eq heterogeneous l3} 
\frac{(s+\mu)(s+\lambda)}{s}\hat{P}_0(x,s+\mu)-\frac{s+\lambda}{s}\delta(x-x_0)=\frac{s+\lambda}{s}\mathcal{D}_\alpha\frac{\partial}{\partial x}\left\{|x|^{\alpha/2}\frac{\partial}{\partial x}\left[|x|^{\alpha/2}\hat{P}_0(x,s+\mu)\right]\right\}.
\end{align}
We notice that if we introduce a new function 
\begin{align}\label{PvsP0}
    \hat{P}(x,s)=\frac{s+\lambda}{s}\hat{P}_0(x,s+\mu),
\end{align}
{}from eq.~(\ref{diffusion eq heterogeneous l3}) we arrive to eq.~(\ref{diffusion eq heterogeneous rm l2}). 
Therefore, by inverse Laplace transform of eq.~(\ref{PvsP0}), we find
\begin{align}\label{renewal appendix}
    P(x,t)=e^{-\mu t}P_0(x,t)+\lambda\int_{0}^{t}e^{-\mu t'}P_0(x,t')\,\D t'\,.
\end{align}
Thus, the solution to the  HDP  with recruitment and mortality can be found simply by knowing the free solution $P_0(x,t)$. We note that in the case where $\lambda=\mu=r$, we obtain the renewal equation for the HDP under Poissonian resetting, i.e.,
\begin{equation}\label{renewal_ap}
    P(x,t)=e^{-r t}P_0(x,t)+r\int_{0}^{t}e^{-r t'}P_0(x,t')\,\D t'\,.
\end{equation}
The PDF of the HDP without recruitment and mortality reads~\cite{jpa_hdp}
\begin{align}\label{PDF str final appendix}
    P_0(x,t)=\frac{|x|^{1/p-1}}{\sqrt{4\pi\mathcal{D}_{\alpha}t}}\exp\left(-\frac{p^{2}\left[\textrm{sgn}(x)\,|x|^{1/p}-\textrm{sgn}(x_0)\,|x_0|^{1/p}\right]^{2}}{4\mathcal{D}_{\alpha}t}\right).
\end{align}

Let us find the norm of the solution. By integration, we have
\begin{align}
    \int_{-\infty}^{\infty}P(x,t)\,\D x\,&=e^{-\mu t}\int_{-\infty}^{\infty}P_0(x,t)\,\D x\,+\lambda\int_{0}^{t}e^{-\mu t'}\int_{-\infty}^{\infty}P_0(x,t')\,\D x\,\D t'\,\nonumber\\&=e^{-\mu t}+\lambda\int_{0}^{t}e^{-\mu t'}\,\D t'\,=1-\frac{\mu-\lambda}{\mu}\left(1-e^{-\mu t}\right)=N(t),
\end{align}
where we use the normalization of the PDF for the HDP, $\int_{-\infty}^{\infty}P_0(x,t)\,\D x\,=1$ (note that for $t\rightarrow\infty$, $N(t)=\lambda/\mu$). This means that the density $P(x,t)$ is not normalized due to the different recruitment and mortality rates. The normalized solution, then becomes
\begin{align}
    P_{N}(x,t)=\frac{1}{N(t)}P(x,t).
\end{align}

{}From here we can calculate the corresponding moments, simply by knowing the moments for the HDP without recruitment and mortality ($\lambda=\mu=0$), i.e.,
\begin{align}
    \langle x^n(t)\rangle&=\int_{-\infty}^{\infty}x^nP_N(x,t)\,\D x\,\nonumber\\&=\frac{e^{-\mu t}}{N(t)}\int_{-\infty}^{\infty}x^nP_0(x,t)\,\D x+\frac{\lambda}{N(t)}\int_{0}^{t}e^{-\mu t'}\int_{-\infty}^{\infty}x^nP_0(x,t')\,\D x\D t'\,
    \nonumber\\
    &=\frac{1}{N(t)}\left[e^{-\mu t}\langle x^n(t)\rangle_0+\lambda\int_0^{t}\!\D t'\:e^{-\mu t'}\langle x^n(t')\rangle_0\right],
\end{align}
where $\langle x^n(t)\rangle_0$ is the $n$-th moment of the HDP without recruitment and mortality, obtained in Ref.~\cite{jpa_hdp}.
For the MSD, we obtain
\begin{align}
    \langle x^2(t)\rangle=\frac{\frac{\Gamma\left(1+2p\right)}{p^{2p}}\frac{\mathcal{D}_\alpha^{p}}{\Gamma\left(1+p\right)}}{N(t)}\Big[&e^{-\mu t}t^p\,{_1}F_{1}\left(-p,\frac{1}{2},-p^{2}\frac{|x_{0}|^{2/p}}{4\mathcal{D}_\alpha t}\right)\nonumber\\&+\lambda\int_0^{t}e^{-\mu t'}{t'}^p\,{_1}F_{1}\left(-p,\frac{1}{2},-p^{2}\frac{|x_{0}|^{2/p}}{4\mathcal{D}_\alpha t'}\right)\D t'\,\Big],
\end{align}
where ${_1}F_{1}\left(a,b,z\right)$ is the generalized hypergeometric series.
For $x_0=0$, it reduces to
\begin{align}
    \langle x^2(t)\rangle&=\frac{1}{N(t)}\frac{\Gamma\left(1+2p\right)}{p^{2p}}\frac{\mathcal{D}_\alpha^{p}}{\Gamma\left(1+p\right)}\left[e^{-\mu t}t^p+\lambda\int_0^{t}e^{-\mu t'}{t'}^p\,\D t'\,\right]\nonumber\\&=\frac{1}{N(t)}\frac{\Gamma\left(1+2p\right)}{p^{2p}}\frac{\mathcal{D}_\alpha^{p}}{\Gamma\left(1+p\right)}\left[e^{-\mu t}t^p+\lambda \frac{\gamma(p+1,\mu t)}{\mu^{p+1}}\right],
\end{align}
where $\gamma(a,z)=\int_{0}^{z}y^{a-1}e^{-y}\,dy$ is the lower incomplete gamma function.

\subsection{Long time asymptotic and stationary state}\label{sec_long}

{}From eq.~(\ref{PvsP0}), in the long time limit we obtain
\begin{align}\label{PvsP0_st}
    P_{st}(x)&=\lim_{t\rightarrow\infty}P(x,t)=\lim_{s\rightarrow0}s\hat{P}(x,s)=\lambda\hat{P}_0(x,\mu)\nonumber\\&=\frac{\lambda|x|^{1/p-1}}{\sqrt{4\mathcal{D}_\alpha\mu}}\exp\Bigg(-p\left|\textrm{sgn}(x)|x|^{1/p}-\textrm{sgn}(x_0)|x_{0}|^{1/p}\right|\sqrt{\frac{\mu}{\mathcal{D}_\alpha}}\Bigg).
\end{align}
This means that the system approaches a stationary state in the long time limit. In the long time limit the MSD saturates to
\begin{align}
    \langle x^2(t)\rangle\sim \frac{\Gamma\left(1+2p\right)}{p^{2p}}\frac{\mathcal{D}_\alpha^{p}}{\mu^{p}}.
\end{align}

\subsection{Transition to the stationary state}

Starting from Eq.~(\ref{renewal_ap}) and using the explicit form of the free heterogeneous diffusion propagator in Eq.~(\ref{PDF str final}), one concludes that the second term is dominant in the long time limit. Let us analyse the second term for large but not infinite time $t$. Thus, we have 
\begin{align}
I(x,t) = \lambda \int_0^t e^{-\mu t'} P_0(x,t')\,\D t'\,= C(x) \int_0^t t'^{-1/2} \exp\left[ -\mu t' - \frac{B(x)}{t'}
\right]\D t'\,,
\end{align}
where
\begin{align}
C(x)
&=
\lambda
\frac{|x|^{1/p-1}}
{\sqrt{4\pi\mathcal{D}_\alpha}},
\\
B(x)
&=
\frac{p^2}{4\mathcal{D}_\alpha}
\left[
\operatorname{sgn}(x)|x|^{1/p}
-
\operatorname{sgn}(x_0)|x_0|^{1/p}
\right]^2 .
\end{align}

In order to analyse the transition to the stationary state, we follow the method of Laplace approximation used in Ref.~\cite{saddle}. We write the integral as, where we introduce $t'=t\tau$, $\D t'=t\D\tau$,
\begin{align}
    I(x,t)=C(x) t^{1/2}\int_0^1 \tau^{-1/2}e^{-t\Phi(t',w)}\D t'\,,
\end{align}
where
\begin{align}
    \Phi(\tau,w)=\mu\tau+\frac{p^2w^2}{4\mathcal{D}_\alpha \tau},
\end{align}
and
\begin{align}
    w=\frac{\left|
\operatorname{sgn}(x)|x|^{1/p}
-
\operatorname{sgn}(x_0)|x_0|^{1/p}\right|}{\tau}.
\end{align}
Then we use the same analysis as below and find the minimum of $\Psi(\tau,w)$, i.e., $\frac{d}{d\tau}\Psi(\tau,w)$ for $\tau\in(0,1)$, and then we calculate $\Phi(\tau,w)$ in $\tau=1$, from where we find the regions in which the steady state is achieved and in which it is in transient regime.

With this we show that $P_r(x)\sim e^{-t\mathcal{I}(w)}$,
where
\begin{align}
\mathcal{I}(w)=\left\lbrace\begin{array}{ll}
   p\sqrt{\frac{\mu}{\mathcal{D}_\alpha}}w,  & \left|
\operatorname{sgn}(x)|x|^{1/p}
-
\operatorname{sgn}(x_0)|x_0|^{1/p}\right|\leq\frac{\sqrt{4\mathcal{D}_\alpha \mu}}{p}t, \\
  \mu+\frac{p^2}{4\mathcal{D}_\alpha}w^2,  & \left|
\operatorname{sgn}(x)|x|^{1/p}
-
\operatorname{sgn}(x_0)|x_0|^{1/p}\right|\geq\frac{\sqrt{4\mathcal{D}_\alpha \mu}}{p}t,
\end{array}\right.
\end{align}
where $\mathcal{I}(w)$ is the large deviation function.

The exponent can be written in terms of the phase function

\begin{align}
\phi(t')=
\mu t'
+
\frac{B(x)}{t'} .
\end{align}

For $x\neq x_0$, one has $B(x)>0$. The stationary point is obtained from
\begin{align}
\frac{d\phi}{\D t'\,}=
\mu-\frac{B(x)}{t'^2}=0,
\end{align}
which gives
\begin{align}
t_*=
\left(\frac{B(x)}{\mu}\right)^{1/2}.
\end{align}
Moreover,
\begin{align}
\phi''(t_*)=
\frac{2B(x)}{t_*^3}>0,
\end{align}
so $t_*$ corresponds to a local minimum of the phase. Therefore, for sufficiently large times such that 
$t_*\in(0,t)$, the Laplace approximation gives
\begin{align}
I(x,t)
\sim
C(x)\,
t_*^{-1/2}
\exp[-\phi(t_*)]
\left(
\frac{2\pi}{\phi''(t_*)}
\right)^{1/2}.
\end{align}
Since
\begin{align}
\phi(t_*)
=
2\sqrt{\mu B(x)},
\end{align}
and
\begin{align}
t_*^{-1/2}
\left(
\frac{2\pi}{\phi''(t_*)}
\right)^{1/2}
=
\frac{\sqrt{\pi}}{\sqrt{\mu}},
\end{align}
we obtain
\begin{align}
I(x,t)
\sim
\lambda
\frac{|x|^{1/p-1}}
{\sqrt{4\mathcal{D}_\alpha\mu}}
\exp\left[
-2\sqrt{\mu B(x)}
\right].
\end{align}

Substituting the definition of $B(x)$ yields
\begin{align}
I(x,t)
\sim
\lambda
\frac{|x|^{1/p-1}}
{\sqrt{4\mathcal{D}_\alpha\mu}}
\exp\left[
-
p
\left|
\operatorname{sgn}(x)|x|^{1/p}
-
\operatorname{sgn}(x_0)|x_0|^{1/p}
\right|
\sqrt{\frac{\mu}{\mathcal{D}_\alpha}}
\right].
\end{align}
Thus, in the long-time limit, the renewal contribution recovers the stationary profile
\begin{align}
P_{st}(x)
=
\lambda
\frac{|x|^{1/p-1}}
{\sqrt{4\mathcal{D}_\alpha\mu}}
\exp\left[
-
p
\left|
\operatorname{sgn}(x)|x|^{1/p}
-
\operatorname{sgn}(x_0)|x_0|^{1/p}
\right|
\sqrt{\frac{\mu}{\mathcal{D}_\alpha}}
\right],
\qquad x\neq x_0 .
\end{align}

The special case $x=x_0$ must be treated separately. In this case $B(x_0)=0$, and therefore the stationary point collapses to the lower integration boundary, $t_*=0$. Hence, the standard Laplace approximation based on an interior minimum is not directly applicable. The renewal contribution becomes
\begin{align}
I(x_0,t)
=
C(x_0)
\int_0^t
t'^{-1/2}e^{-\mu t'}\D t'\,,
\end{align}
which can be written as
\begin{align}
I(x_0,t)
=
C(x_0)
\frac{\gamma\left(1/2,\mu t\right)}
{\mu^{1/2}} .
\end{align}
Taking the limit $t\to\infty$ and using
\begin{align}
\gamma\left(1/2,\mu t\right)
\to
\Gamma(1/2)=\sqrt{\pi},
\end{align}
one obtains
\begin{align}
P_{st}(x_0)
=
\frac{\lambda |x_0|^{1/p-1}}
{\sqrt{4\mathcal{D}_\alpha\mu}},
\end{align}
which is consistent with the general stationary expression evaluated at $x=x_0$.

This analysis shows that the approach to the stationary state is controlled by the competition between the mortality rate and the effective heterogeneous diffusion distance. In particular, the stationary profile has a large-deviation-like form
\begin{align}
P_{st}(x)\sim \exp[-\mathcal{I}(x)],
\end{align}
with rate function
\begin{align}
\mathcal{I}(x)
=
p
\left|
\operatorname{sgn}(x)|x|^{1/p}
-
\operatorname{sgn}(x_0)|x_0|^{1/p}
\right|
\sqrt{\frac{\mu}{\mathcal{D}_\alpha}} .
\end{align}
Starting from Eq.~(19), 
and using the asymptotic relation 
$\gamma(p+1,\mu t)\to \Gamma(p+1)$ as $t\to\infty$, the stationary MSD becomes
\begin{align}
    \langle x^2\rangle_{st}
    =
    \frac{\Gamma(1+2p)\,\mathcal{D}_\alpha^p}
    {\mu^p\,p^{2p}}.
\end{align}
\section{Calculation of the first passage time}\label{appb}
To compute the underlying first-passage time density $f_0(t)$, we start from the free HDP propagator in Eq.~(\ref{PDF str final}). This propagator can be written in terms of the effective coordinate~\cite{jpa_hdp}
\begin{equation}
    y=
    p\,\operatorname{sgn}(x)|x|^{1/p},
    \qquad
    p=\frac{2}{2-\alpha}.
\end{equation}
In this coordinate, the heterogeneous diffusion process is mapped into ordinary Brownian diffusion with diffusion coefficient $\mathcal{D}_\alpha$. Therefore, the free propagator becomes
\begin{equation}
    \Pi_0(y,t|y_0)
    =
    \frac{1}{\sqrt{4\pi \mathcal{D}_\alpha t}}
    \exp\left[
    -\frac{(y-y_0)^2}{4\mathcal{D}_\alpha t}
    \right],
\end{equation}
where
\begin{equation}
    y_0=
    p\,\operatorname{sgn}(x_0)|x_0|^{1/p}.
\end{equation}

Let the absorbing target be located at $x=L$. In the effective coordinate, the absorbing boundary is located at
\begin{equation}
    y_L=
    p\,\operatorname{sgn}(L)|L|^{1/p}.
\end{equation}
For definiteness, we assume $y_0<y_L$, so that the trajectory starts to the left of the absorbing boundary. The case $y_0>y_L$ is obtained analogously by replacing $y_L-y_0$ with $|y_L-y_0|$. The absorbing condition is then
\begin{equation}
    \Pi_0^{(a)}(y_L,t|y_0)=0,
\end{equation}
which is not satisfied by the free propagator. Therefore, we construct the absorbing propagator by the method of images. The image point associated with $y_0$ is
\begin{equation}
    y_{\mathrm{im}}=2y_L-y_0,
\end{equation}
thus, the absorbing propagator is
\begin{align}
    \Pi_0^{(a)}(y,t|y_0)
    &=
    \Pi_0(y,t|y_0)
    -
    \Pi_0(y,t|y_{\mathrm{im}})
    \nonumber\\
    &=
    \frac{1}{\sqrt{4\pi \mathcal{D}_\alpha t}}
    \left[
    \exp\left(
    -\frac{(y-y_0)^2}{4\mathcal{D}_\alpha t}
    \right)
    -
    \exp\left(
    -\frac{(y-(2y_L-y_0))^2}{4\mathcal{D}_\alpha t}
    \right)
    \right].
    \label{absorbing_propagator_y}
\end{align}
Indeed, evaluating Eq.~\eqref{absorbing_propagator_y} at $y=y_L$ gives
\begin{align}
    \Pi_0^{(a)}(y_L,t|y_0)
    =
    \frac{1}{\sqrt{4\pi \mathcal{D}_\alpha t}}
    \left[
    \exp\left(
    -\frac{(y_L-y_0)^2}{4\mathcal{D}_\alpha t}
    \right)
    -
    \exp\left(
    -\frac{(y_L-(2y_L-y_0))^2}{4\mathcal{D}_\alpha t}
    \right)
    \right] =
    0.
\end{align}
The survival probability of the underlying HDP, without mortality and recruitment, is obtained by integrating the absorbing propagator over the accessible domain,
\begin{equation}
    S_0(t)
    =
    \int_{-\infty}^{y_L}
    \Pi_0^{(a)}(y,t|y_0)\,dy.
\end{equation}
Using Eq.~\eqref{absorbing_propagator_y}, this gives
\begin{align}
    S_0(t)
    =
    \int_{-\infty}^{y_L}
    \frac{1}{\sqrt{4\pi \mathcal{D}_\alpha t}}
    \exp\left[
    -\frac{(y-y_0)^2}{4\mathcal{D}_\alpha t}
    \right]dy -
    \int_{-\infty}^{y_L}
    \frac{1}{\sqrt{4\pi \mathcal{D}_\alpha t}}
    \exp\left[
    -\frac{(y-(2y_L-y_0))^2}{4\mathcal{D}_\alpha t}
    \right]dy .
\end{align}
Introducing
\begin{equation}
    a=
    \frac{y_L-y_0}{\sqrt{4\mathcal{D}_\alpha t}},
\end{equation}
the two Gaussian integrals become
\begin{equation}
    \frac{1}{2}\left[1+\operatorname{erf}(a)\right],
    \qquad
    \frac{1}{2}\left[1-\operatorname{erf}(a)\right],
\end{equation}
respectively. Therefore,
\begin{equation*}
    S_0(t)
    =
    \operatorname{erf}
    \left(
    \frac{y_L-y_0}{\sqrt{4\mathcal{D}_\alpha t}}
    \right).
\end{equation*}
Equivalently, by defining the effective distance to the target as
\begin{equation}
    \Delta y=|y_L-y_0|,
\end{equation}
we write
\begin{equation}
    S_0(t)
    =
    \operatorname{erf}
    \left(
    \frac{\Delta y}{\sqrt{4\mathcal{D}_\alpha t}}
    \right).
    \label{HDP_base_survival}
\end{equation}
The first-passage time density is then obtained from
\begin{equation}
    f_0(t)
    =
    -\frac{\D S_0(t)}{\D t}.
\end{equation}
Using
\begin{equation}
    \frac{d}{dz}\operatorname{erf}(z)
    =
    \frac{2}{\sqrt{\pi}}e^{-z^2},
\end{equation}
with $z(t)=
    \frac{\Delta y}{\sqrt{4\mathcal{D}_\alpha t}}$,
we obtain
\begin{align}
    f_0(t)
    =
    -\frac{2}{\sqrt{\pi}}
    e^{-z(t)^2}
    \frac{\D z}{\D t} =
    \frac{\Delta y}
    {\sqrt{4\pi\mathcal{D}_\alpha t^3}}
    \exp\left[
    -\frac{\Delta y^2}{4\mathcal{D}_\alpha t}
    \right].
    \label{HDP_base_FPT_density}
\end{align}
Thus, the first-passage time density of the underlying HDP has the same form as the Brownian first-passage density in the effective coordinate $y$, with the physical distance replaced by the heterogeneous diffusion distance $\Delta y=|y_L-y_0|$. Although the first-passage calculation is most conveniently performed in the effective coordinate $y$, the physical dependence on the original spatial coordinate is recovered through the effective distance
\begin{equation}
\Delta y (x_0,L,\alpha)
=
p\left|
\operatorname{sgn}(L)|L|^{1/p}
-
\operatorname{sgn}(x_0)|x_0|^{1/p}
\right|
=
\frac{2}{2-\alpha} \left|
\operatorname{sgn}(L)|L|^{\frac{2-\alpha}{2}}
-
\operatorname{sgn}(x_0)|x_0|^{\frac{2-\alpha}{2}}
\right|.
\label{effective_distance_physical}
\end{equation}
Therefore, all first-passage quantities depend on the original target position $L$, the initial position $x_0$, and the heterogeneity exponent $\alpha$ through $\Delta y(x_0,L,\alpha)$.
This dependence is physically relevant because the process does not measure the Euclidean distance $|L-x_0|$ alone, but rather the distance induced by the spatially dependent diffusivity. For the simple case $0<x_0<L$, Eq.~\eqref{effective_distance_physical} reduces to
\begin{equation}
    \Delta y
    =
    \frac{2}{2-\alpha}
    \left(
    L^{(2-\alpha)/2}
    -
    x_0^{(2-\alpha)/2}
    \right).
    \label{effective_distance_positive_axis_ap}
\end{equation}
For $\alpha=0$, one recovers the homogeneous Brownian result $\Delta y=L-x_0$. For $\alpha>0$, the diffusivity $\mathcal{D}(x)=\mathcal{D}_\alpha |x|^\alpha$ increases with $|x|$, so the effective distance grows sublinearly with the target position and the target becomes dynamically closer. Conversely, for $\alpha<0$, the diffusivity decreases with $|x|$, the effective distance grows superlinearly, and the target becomes dynamically farther.

\noindent We now evaluate the survival probability of a single mortal trajectory, without recruitment. From Eq.~\eqref{FPT_single_mortal_survival} and using Eq.~\eqref{HDP_base_FPT_density}, we obtain

\begin{align}
    S_{\mu}(t)
    &=
    1-
    \int_0^t
    e^{-\mu t'}
    \frac{\Delta y}
    {\sqrt{4\pi \mathcal{D}_\alpha t'^3}}
    \exp\left[
    -\frac{\Delta y^2}
    {4\mathcal{D}_\alpha t'}
    \right]\D t'\,
    \nonumber\\
    &=
    1-
    \frac{\Delta y}
    {\sqrt{4\pi \mathcal{D}_\alpha}}
    \int_0^t
    t'^{-3/2}
    \exp\left[
    -\mu t'
    -
    \frac{\Delta y^2}
    {4\mathcal{D}_\alpha t'}
    \right]\D t'\,.
    \label{single_mortal_survival_integral}
\end{align}

The remaining integral belongs to the standard class
\begin{equation}
    \int t^{-3/2}
    \exp\left[-\left(\frac{a^2}{t}+b^2t\right)\right]\D t\,,
\end{equation}
whose antiderivatives are naturally expressed in terms of complementary error functions. Introducing
\begin{equation}
    a=
    \frac{\Delta y}{\sqrt{4\mathcal{D}_\alpha}},
    \qquad
    b=\sqrt{\mu},
\end{equation}
and using the complementary error function,

\begin{equation}
    \operatorname{erfc}(z)
    =
    \frac{2}{\sqrt{\pi}}
    \int_z^\infty e^{-u^2}\,du,
\end{equation}
this gives
\begin{align}
    \int_0^t
    e^{-\mu t'}f_0(t')\,\D t'\,
    =
    \frac{1}{2}
    \exp\left[
    -\Delta y\sqrt{\frac{\mu}{\mathcal{D}_\alpha}}
    \right]
    \operatorname{erfc}
    \left(
    \frac{\Delta y}{\sqrt{4\mathcal{D}_\alpha t}}
    -
    \sqrt{\mu t}
    \right) +
    \frac{1}{2}
    \exp\left[
    \Delta y\sqrt{\frac{\mu}{\mathcal{D}_\alpha}}
    \right]
    \operatorname{erfc}
    \left(
    \frac{\Delta y}{\sqrt{4\mathcal{D}_\alpha t}}
    +
    \sqrt{\mu t}
    \right).
    \label{mortal_arrival_probability}
\end{align}
Therefore, the survival probability of a single mortal HDP trajectory is
\begin{align}
    S_{\mu}(t)
    =
    1
    -
    \frac{1}{2}
    \exp\left[
    -\Delta y\sqrt{\frac{\mu}{\mathcal{D}_\alpha}}
    \right]
    \operatorname{erfc}
    \left(
    \frac{\Delta y}{\sqrt{4\mathcal{D}_\alpha t}}
    -
    \sqrt{\mu t}
    \right) -
    \frac{1}{2}
    \exp\left[
    \Delta y\sqrt{\frac{\mu}{\mathcal{D}_\alpha}}
    \right]
    \operatorname{erfc}
    \left(
    \frac{\Delta y}{\sqrt{4\mathcal{D}_\alpha t}}
    +
    \sqrt{\mu t}
    \right).
    \label{append single_mortal_survival_exact_ap}
\end{align}

This expression has two useful limiting cases. First, in the absence of mortality, $\mu\to0$, Eq.~\eqref{append single_mortal_survival_exact_ap} reduces to the survival probability of the underlying HDP (Eq.~\ref{HDP_base_survival}). Second, in the long-time limit,
\begin{equation}
    \lim_{t\to\infty}
    \int_0^t
    e^{-\mu t'}f_0(t')\,\D t'\,
    =
    \exp\left[
    -\Delta y\sqrt{\frac{\mu}{\mathcal{D}_\alpha}}
    \right],
\end{equation}
whence,
\begin{equation}
    S_{\mu}(\infty)
    =
    1-
    \exp\left[
    -\Delta y\sqrt{\frac{\mu}{\mathcal{D}_\alpha}}
    \right].
\end{equation}

It is worth noting that the same limiting result can be recovered by applying Laplace's method to the integral in Eq.~\eqref{single_mortal_survival_integral}. In this case, the dominant contribution is controlled by the minimum of the phase
\begin{equation}
    \Phi(t')
    =
    \mu t'
    +
    \frac{\Delta y^2}{4\mathcal{D}_\alpha t'} ,
\end{equation}
which occurs at
\begin{equation}
    t_*
    =
    \frac{\Delta y}{2\sqrt{\mathcal{D}_\alpha\mu}} .
\end{equation}
The corresponding value of the phase is
\begin{equation}
    \Phi(t_*)
    =
    \Delta y
    \sqrt{\frac{\mu}{\mathcal{D}_\alpha}} .
\end{equation}
Therefore, for observation times sufficiently larger than $t_*$, Laplace's method gives
\begin{equation}
    \int_0^t
    e^{-\mu t'}f_0(t')\,\D t'\,
    \simeq
    \exp\left[
    -\Delta y
    \sqrt{\frac{\mu}{\mathcal{D}_\alpha}}
    \right]
    =
    \mathcal{P}_{\mu}.
\end{equation}
Thus, the Laplace approximation reproduces the exact result only in the stationary, long-time limit of the single mortal survival probability,
\begin{equation}
    S_{\mu}(t)
    \simeq
    1-\mathcal{P}_{\mu},
    \qquad
    t\gg t_*.
\end{equation}
Therefore, it should not be interpreted as an approximation to the full finite-time expression of $S_{\mu}(t)$, which is instead given by Eq.~\eqref{survival_exact}.
The probability that a single mortal trajectory reaches the target before being removed is
\begin{equation}
    \mathcal{P}_{\mu}
    =
    \mathbb{P}(\tau<T_m)
    =
    \int_0^\infty
    e^{-\mu t}f_0(t)\,\D t\,
    =
    \exp\left[
    -\Delta y\sqrt{\frac{\mu}{\mathcal{D}_\alpha}}
    \right],
    \label{single_mortal_success_probability_HDP}
\end{equation}
where $T_m$ denotes the exponentially distributed mortality time with rate $\mu$.

The numerical evaluation was implemented in Python. The complementary error functions were evaluated using \texttt{scipy.special}, the cumulative time integrals were computed using the trapezoidal rule through \texttt{scipy.integrate.cumulative\_trapezoid}, and the mean first-passage times were obtained with \texttt{scipy.integrate.trapezoid}.
For a discrete time grid $0=t_0<t_1<\cdots<t_N=t$, the cumulative recruitment contribution is approximated as
\begin{equation}
    I_{\mu}(t_n)
    =
    \int_0^{t_n}
    \left[
    1-S_{\mu}(t')
    \right]\D t'\,
    \simeq
    \sum_{j=1}^{n}
    \frac{
    \left[1-S_{\mu}(t_j)\right]
    +
    \left[1-S_{\mu}(t_{j-1})\right]
    }{2}
    \Delta t_j ,
    \label{recruitment_integral_trapezoid_ap}
\end{equation}
where $\Delta t_j=t_j-t_{j-1}$. Thus,
\begin{equation}
    S=
    S_{\mu}(t_n)
    \exp\left[
    -\lambda I_{\mu}(t_n)
    \right].
    \label{full_survival_numerical_ap}
\end{equation}

The corresponding first-passage time density can then be obtained from
numerical differentiation of 
Eq.~\eqref{full_survival_numerical_ap} and their results are shown in Fig.~\ref{fig_FPT_recruitment_mortality} of the main text. To illustrate explicitly how recruitment reshapes 
$S(t)$ at fixed mortality rate, Fig. B1 shows the same numerical evaluation of Eq.\eqref{full_survival_numerical_ap}, swept over the recruitment rate $\lambda$.

\begin{figure}[htbp]
    \centering
    \includegraphics[width=0.55\textwidth]{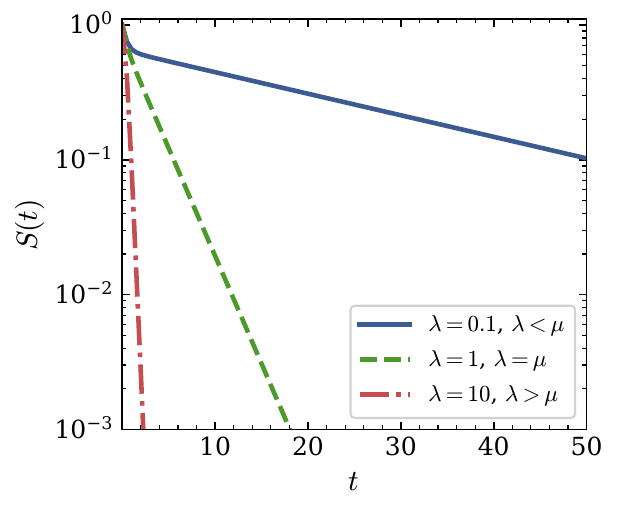}
    \caption{
    Effect of the recruitment rate $\lambda$ on the full recruitment--mortality survival probability $S(t)$, for fixed mortality rate $\mu=1.0$, in dimensionless units.
    The three curves correspond to representative regimes with $\lambda<\mu$, $\lambda=\mu$, and $\lambda>\mu$.
    Increasing $\lambda$ accelerates the decay of $S_{\lambda,\mu}(t)$ because each recruited trajectory provides an additional independent attempt to reach the target.
    Parameters: $\Delta y=1$ and $\mathcal{D}_\alpha=1$.
    }
    \label{fig_recruitment_survival_appendix}
\end{figure}
\begin{figure}[htbp]
    \centering
    \includegraphics[width=0.6\textwidth]{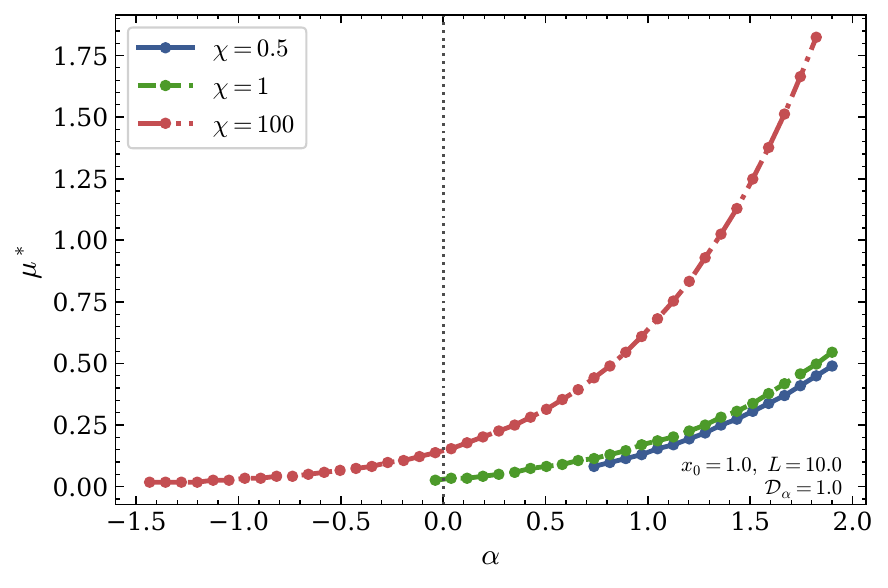}
    \caption{
    Optimal mortality rate $\mu^*$ as a function of the heterogeneity exponent $\alpha$, for selected values of the recruitment-to-mortality ratio $\chi$, in dimensionless units.
    For each value of $\alpha$, the effective distance $\Delta y(x_0,L,\alpha)$ is computed for fixed $x_0=1$ and $L=10$, and $\mathcal{D}_\alpha=1$. The optimal mortality rate is obtained by minimizing the numerical mean first-passage time over the mortality parameter $\mu$.
    }
    \label{fig_optimal_mortality_vs_alpha_appendix}
\end{figure}

As shown in Fig.~\ref{fig_optimal_mortality_vs_alpha_appendix}, spatial heterogeneity modifies the effective distance $\Delta y(x_0,L,\alpha)$ and therefore shifts the optimal mortality rate $\mu^*$. The optimal value increases with $\alpha$, consistent with the reduction of the effective distance as the target becomes dynamically closer. We also observe that, for recruitment--mortality ratios below unity, $\chi<1$, internal optima are not found for the whole range of $\alpha$ considered. This suggests that when recruitment is weak relative to mortality, the existence of a well-defined optimal mortality rate becomes more sensitive to the heterogeneity of the medium.

\newpage

{\small 
\begin{thebibliography}{999}

\bibitem{00}
L. A. Rastrigin, ``Extremal control by the method of random scanning,'' Avtom. Telemekh. \textbf{21}, 1201 (1960).

\bibitem{01}
W. Wang, A. G. Cherstvy, X. Liu, and R. Metzler, ``Anomalous diffusion and nonergodicity for heterogeneous diffusion processes with fractional Gaussian noise,'' Phys. Rev. E \textbf{102}, 012146 (2020).

\bibitem{02}
Y. Xu, X. Liu, Y. Li, and R. Metzler, ``Heterogeneous diffusion processes and nonergodicity with Gaussian colored noise in layered diffusivity landscapes,'' Phys. Rev. E \textbf{102}, 062106 (2020).

\bibitem{03}
J. W. Haus and K. W. Kehr, ``Diffusion in regular and disordered lattices,'' Phys. Rep. \textbf{150}, 263 (1987).

\bibitem{04}
J.-P. Bouchaud and A. Georges, ``Anomalous diffusion in disordered media: Statistical mechanisms, models and physical applications,'' Phys. Rep. \textbf{195}, 127 (1990).

\bibitem{05}
S. Havlin and D. Ben-Avraham, ``Diffusion in disordered media,'' Adv. Phys. \textbf{51}, 187 (2002).

\bibitem{06}
R. Metzler and J. Klafter, ``The random walk's guide to anomalous diffusion: A fractional dynamics approach,'' Phys. Rep. \textbf{339}, 1 (2000).

\bibitem{report}
R. Metzler and J. Klafter, ``The restaurant at the end of the random walk: Recent developments in the description of anomalous transport by fractional dynamics,'' J. Phys. A \textbf{37}, R161 (2004).

\bibitem{08}
S. Burov, J.-H. Jeon, R. Metzler, and E. Barkai, ``Single particle tracking in systems showing anomalous diffusion: The role of weak ergodicity breaking,'' Phys. Chem. Chem. Phys. \textbf{13}, 1800 (2011).

\bibitem{09}
E. Barkai, Y. Garini, and R. Metzler, ``Strange kinetics of single molecules in living cells,'' Phys. Today \textbf{65}(8), 29 (2012).

\bibitem{010}
I. M. Sokolov, ``Models of anomalous diffusion in crowded environments,'' Soft Matter \textbf{8}, 9043 (2012).

\bibitem{011}
F. H\"ofling and T. Franosch, ``Anomalous transport in the crowded world of biological cells,'' Rep. Prog. Phys. \textbf{76}, 046602 (2013).

\bibitem{012}
Y. Meroz and I. M. Sokolov, ``A toolbox for determining subdiffusive mechanisms,'' Phys. Rep. \textbf{573}, 1 (2015).

\bibitem{013}
R. Metzler, J.-H. Jeon, and A. G. Cherstvy, ``Non-Brownian diffusion in lipid membranes: Experiments and simulations,'' Biochim. Biophys. Acta Biomembr. \textbf{1858}, 2451 (2016).

\bibitem{app1}
F. Gjetvaj, A. Russian, P. Gouze, and M. Dentz, ``Dual control of flow field heterogeneity and immobile porosity on non-Fickian transport in Berea sandstone,'' Water Resour. Res. \textbf{51}, 8273 (2015).

\bibitem{app6}
B. O'Shaughnessy and I. Procaccia, ``Analytical solutions for diffusion on fractal objects,'' Phys. Rev. Lett. \textbf{54}, 455 (1985).

\bibitem{app2}
O. G. Bakunin, ``Correlation effects and turbulent diffusion scalings,'' Rep. Prog. Phys. \textbf{67}, 965 (2004).

\bibitem{app3}
B. Sawford, ``Turbulent relative dispersion,'' Annu. Rev. Fluid Mech. \textbf{33}, 289 (2001).

\bibitem{app4}
C. Castellano, S. Fortunato, and V. Loreto, ``Statistical physics of social dynamics,'' Rev. Mod. Phys. \textbf{81}, 591 (2009).

\bibitem{app5}
R. Kazakevi\v{c}ius and A. Kononovi\v{c}ius, ``Anomalous diffusion in nonlinear transformations of the noisy voter model,'' Phys. Rev. E \textbf{103}, 032154 (2021).

\bibitem{report2}
R. Metzler, J.-H. Jeon, A. G. Cherstvy, and E. Barkai, ``Anomalous diffusion models and their properties: Non-stationarity, non-ergodicity, and ageing at the centenary of single particle tracking,'' Phys. Chem. Chem. Phys. \textbf{16}, 24128 (2014).

\bibitem{chaos_rev}
T. Sandev, L. Kocarev, and R. Metzler, ``Anomalous diffusion and fluctuations in complex systems and networks,'' Chaos \textbf{36}, 010401 (2026).

\bibitem{Leibovich}
N. Leibovich and E. Barkai, ``Infinite ergodic theory for heterogeneous diffusion processes,'' Phys. Rev. E \textbf{99}, 042138 (2019).

\bibitem{1}
K. It\^{o}, ``Stochastic integral,'' Proc. Imp. Acad. Tokyo \textbf{20}, 519 (1944).

\bibitem{2}
I. I. Gikhman and A. V. Skorokhod, \textit{The Theory of Stochastic Processes III} (Springer, Berlin, 1979).

\bibitem{stratonovich}
R. L. Stratonovich, ``A new representation for stochastic integrals and equations,'' SIAM J. Control \textbf{4}, 362 (1966).

\bibitem{4}
P. H\"anggi, ``Stochastic processes I: Asymptotic behaviour and symmetries,'' Helv. Phys. Acta \textbf{51}, 183 (1978).

\bibitem{5}
Y. L. Klimontovich, `Ito, Stratonovich and kinetic forms of stochastic equations,'' Physica A \textbf{163}, 515 (1990); Y. L. Klimontovich, `Nonlinear Brownian motion,'' Phys.-Usp. \textbf{37}, 737 (1994).

\bibitem{chchme}
A. G. Cherstvy, A. V. Chechkin, and R. Metzler, ``Anomalous diffusion and ergodicity breaking in heterogeneous diffusion processes,'' New J. Phys. \textbf{15}, 083039 (2013).

\bibitem{jpa_hdp}
T. Sandev, V. Domazetoski, L. Kocarev, R. Metzler, and A. Chechkin, ``Heterogeneous diffusion with stochastic resetting,'' J. Phys. A: Math. Theor. \textbf{55}, 074003 (2022).

\bibitem{evans_maj}
M. R. Evans and S. N. Majumdar, ``Diffusion with stochastic resetting,'' Phys. Rev. Lett. \textbf{106}, 160601 (2011).

\bibitem{evans_rev}
M. R. Evans, S. N. Majumdar, and G. Schehr, ``Stochastic resetting and applications,'' J. Phys. A: Math. Theor. \textbf{53}, 193001 (2020).

\bibitem{linn2026dynamic}
S. Linn and A. Kumar, ``Dynamic redundancy and mortality in stochastic search,'' Phys. Rev. Lett. \textbf{137}, 097102 (2026).

\bibitem{mercado2026stochastic}
G. Mercado-V\'asquez and D. Boyer, ``Stochastic synthesis-degradation processes: First-passage properties and connections with resetting,'' Phys. Rev. E \textbf{114}, L022107 (2026).

\bibitem{arxiv_gbm}
S. Pal, V. Stojkoski, A. Pal, and T. Sandev, ``Geometric Brownian motion with intermittent entries and exits,'' arXiv:2605.17299 (2026).

\bibitem{Forgacs2022}
P. Forg\'acs, A. Lib\'al, C. Reichhardt, N. Hengartner, and C. J. O. Reichhardt, ``Using active matter to introduce spatial heterogeneity to the susceptible infected recovered model of epidemic spreading,'' Sci. Rep. \textbf{12}, 11229 (2022).

\bibitem{Chindra2025}
A. Chinprateep, ``Post-COVID-19 epidemiological insights and SEIR modeling for future global pandemic preparedness: Lessons from the Western Pacific,'' medRxiv (2025), 10.1101/2025.08.12.25333519.

\bibitem{Siegenfeld2022}
A. F. Siegenfeld, P. K. Kollepara, and Y. Bar-Yam, ``Modeling complex systems: A case study of compartmental models in epidemiology,'' Complexity, 3007864 (2022).

\bibitem{Pokhrel2026}
S. Pokhrel, N. Sharma, and R. R. B. S. Thakuri, ``Mathematical models and their applications in understanding the dynamics of infectious diseases,'' J. Appl. Math., 3007179 (2026).

\bibitem{Bobashev2008}
G. Bobashev, R. J. Morris, D. M. Goedecke. ``Sampling for global epidemic models and the topology of an international airport network,'' PLoS One \textbf{3.(9)}, e3154 (2008). 

\bibitem{Meerson2015}
B. Meerson and S. Redner, ``Mortality, redundancy, and diversity in stochastic search,'' Phys. Rev. Lett. \textbf{114}, 198101 (2015).

\bibitem{Campos2024}
D. Campos and V. M\'endez, ``Dynamic redundancy as a mechanism to optimize collective random searches,'' Phys. Rev. E \textbf{109}, 064109 (2024).

\bibitem{saddle}
S. N. Majumdar, S. Sabhapandit, and G. Schehr, ``Dynamical transition in the temporal relaxation of stochastic processes under resetting,'' Phys. Rev. E \textbf{91}, 052131 (2015).

\end{thebibliography}
}
\end{document}